\documentclass[11pt]{article}

\usepackage{amsmath,amssymb,graphicx}

\renewcommand{\title}[1]{%
	\begin{center} \Large \bf #1 \end{center}%
	}
\renewcommand{\author}[2]{%
	\begin{center} {\it #1}  \vspace{2mm}\\ %
	  #2%
	\end{center}%
	\addvspace{\baselineskip}%
	}

\begin{document}
\newpage
\setcounter{section}{0}
\setcounter{equation}{0}
\setcounter{figure}{0}
\baselineskip 5mm
\title{%
Can Family Gauge Bosons Be Visible\\
 by Terrestrial Experiments? \footnote{
Invited talk at CST-MISC Joint Symposium on Particle Physics 
--- from Spacetime Dynamics to Phenomenology --- \\
Published in JPS Conf. Proc. {\bf 7}, 010009 (2015)}
}  %
\author{%
Department of Physics, Osaka University 
}{%
Yoshio Koide 
}%
\begin{quotation}
It is investigated whether observations of family gauge bosons by terrestrial 
experiments are possible or not.  We propose an extended version of 
Sumino's family gauge boson model based on U(3) family symmetry. 
Then, we can expect the lowest family gauge boson $A_1^1$ with $M \sim
4.3$ TeV.
\end{quotation}



\section{Why not consider a family gauge symmetry? }

{\bf 1.1 Basic standpoints}

A degree of freedom of ``families" is the last one
 which has still not been accepted as a  gauge symmetry in the standard model (SM).  
The idea of family gauge bosons is the most natural extension of SM.
If the family gauge symmetry is absent, the Cabibbo-Kobayashi-Maskawa (CKM) mixing 
$V_{CKM} = U_u^\dagger U_d$ is observable,  while the quark mixing matrices 
$U_u$ and $U_d$ cannot be observable! 
I think that a theory which includes such unobservable quantities is incomplete. 
 
If family gauge bosons really exist, I believe that those should be particles 
which can be observed by terrestrial experiments.

\vspace{2mm}
\noindent
{\bf 1.2 Background  knowledge: Sumino  model and its extended one}

The present model is highly affected by the Sumino model \cite{Sumino_PLB09}
and an extended version (the K-Y model) \cite{K-Y_PLB12}.
Therefore, first, let us give a brief review of the Sumino model and K-Y model.



In 2009, Sumino \cite{Sumino_PLB09} has seriously taken why the mass formula
\cite{Koide-mass} 
$$
K \equiv \frac{m_e + m_\mu +m_\tau}{ \left(\sqrt{m_e}+ 
\sqrt{m_\tau} +\sqrt{m_\tau}\right)^2} = \frac{2}{3}, 
\eqno(1)
$$                                                                
is so remarkably satisfied with  the pole masses, while if we 
 take the running masses, the agreement is somewhat spoiled.  
The deviation is caused by a factor $\log(m_e^2/\mu^2)$ in 
the QED radiative correction.  
If the logarithmic term is absent, the formula can be invariant 
under the running masses.
Therefore, Sumino has assumed that there are family gauge bosons
 whose masses are proportional to the charged lepton masses, i.e.
$M_{ij} \propto m_{ei}$, and thereby, the factor $\log(m_{ei}^2/\mu^2)$ 
in the QED radiative correction is canceled by a factor 
$\log(M_{ii}^2/\mu^2)$ in the family gauge boson contribution. 

In the K-Y model, 
in order to cancel the factor $\log(m_{ei}^2/ \mu^2)$ in the CED contribution
 by the family gauge boson contribution $\log(M_{ii}^2/ \mu^2)$, the gauge boson
masses take an inverted mass hierarchy because only $M_{ii}^2 =k/m_{ei}$ can 
naturally provide a minus 
sign as $\log M_{ii}^2 = - \frac{1}{2} \log m_{ei}^2 + \log k$.

The model has the following characteristics: 
(i) Family gauge bosons exist in the mass-eigenstates on the basis in which 
the charged lepton mass matrix is diagonal.  
(ii) Up- and down quark mass matrices are, in general, not diagonal, 
so that the family gauge boson interactions are given by
$$
{\cal H}_{fam} = \frac{g_{F}}{\sqrt{2}} \left[ (\bar{e}_i \gamma_\mu e_j) 
+ (\bar{\nu}_i \gamma_\mu \nu_j) 
+  U^{d*}_{ik} U_{jl}^d(\bar{d}_k \gamma_\mu d_l) 
+ U^{u*}_{ik} U_{jl}^u (\bar{u}_k \gamma_\mu u_l ) \right] (A_i^{\ j})^\mu ,
\eqno(2)
$$
where $U^u$ and $U^d$ are quark mixing matrices. 
That is, family-number violation is caused only in the quark sector.
(iii)  The family symmetry U(3) is broken by a scalar with
$({\bf 3},{\bf 3}^*)$ of U(3)$\times$U(3)$'$, so that the direct 
transition $A_i^{\ j} \leftrightarrow A_j^{\ i}$ cannot appear, so that
the family gauge boson interactions are given only by Eq.(2).


\section{Phenomenology of family gauge bosons with lower masses}

\noindent
{\bf 2.1   Rare decays of ps-mesons}

First, we investigate lower limits of the family gauge boson masses from 
the present data \cite{PDG12} of rare $K$ and $B$ decay searches. 
Here, we denote the lower masses by effective masses $\tilde{M}_{ij} \equiv 
{M_{ij}}/{(g_F/\sqrt{2})}$,
because $\tilde{M}_{ij}$  is useful in discussing the four fermion interactions 
of quarks and leptons rather than the real masses $M_{ij}$.  

Roughly speaking, we conclude that $\tilde{M}_{12} \geq 250$ TeV 
and $\tilde{M}_{23} \geq 7$ TeV.
This is in favor of the inverted mass picture of $A_i^{\ j}$.



Only $Br(K^+ \rightarrow \pi^+ \nu \bar{\nu})$
has been reported with a finite value of the branching ratio,
$(1.7 \pm 1.1) \times 10^{-10}$ \cite{PDG12}. 
It is usually taken that this value is consistent with
the standard model prediction 
\cite{B-Kpinunu} $Br(K^+ \rightarrow \pi^+ \nu \bar{\nu})_{SM} = 
(0.80 \pm 0.11) \times 10^{-10}$.
Since our purpose is to find a room for new physics as much as possible,
we take the center value of the observed value.
Then, we can obtain a value $\tilde{M}_{12} \sim 243$ TeV.

\vspace{2mm}
\noindent
{\bf 2.2 $\mu$-$e$ conversion}

If we consider $\tilde{M}_{12} \sim 250$ TeV, we can obtain rough 
estimate $R_q \equiv \sigma(\mu^- + q \rightarrow e^- + q)/\sigma(\mu^- + u 
\rightarrow \nu_\mu + d) \sim 10^{-14}$. 
(For details, see Ref.\cite{harmless}).
The estimated values $R_d \sim 10^{-14}$ become within 
reach of our observation. 
In fact, an experiment with $R \sim 10^{-17}$ is planed by the
COMET group \cite{COMET}.
(Note that the estimated value $R_q$ has different physical meaning  
from an observed value $R_N \equiv \sigma(\mu^- + (A,Z) \rightarrow 
e^- + (A,Z))/\sigma(\mu^- + (A,Z) \rightarrow \nu_\mu + (A,Z-1))$, 
we consider that the order of the value $R_q$
can provide one with useful information.)

Since the decay $\mu^- \rightarrow e^- +\gamma$ is highly 
suppressed in the present model, if we observe 
$\mu^- N \rightarrow e^- N$ in spite of no observation of 
$\mu^- \rightarrow e^- +\gamma$, 
then it will strongly support our family gauge boson scenario.

\vspace{2mm}
\noindent
{\bf 2.3  Direct production of $A_3^{\ 3}$ and $A_2^{\ 3}$} 

If $A_3^{\ 3}$ and $A_2^{\ 3}$ are light family gauge bosons,
we have a possibility of direct observations of those bosons, i.e. 
$pp \rightarrow A_3^{\ 3} + b +\bar{b} + X \rightarrow 
\tau^- \tau^+ + X$ and $pp \rightarrow A_2^{\ 3} + b +\bar{s} + X 
\rightarrow \mu^- \tau^+ + X$, 
with the branching ratios $Br(A_i^{\ j} \rightarrow \ell_i \bar{\ell}_j)
= 2/15=13.3$ \%.
(Note that the values $\tilde{M}_{ij}$ are effective masses,
values of the real masses $M_{ij}$ are considerably lower than 
$\tilde{M}_{ij}$.)

Note that a value of the branching ratio 
$Br(A_i^{\ j} \rightarrow \nu_i \bar{\nu}_j)$ is 
different according as the neutrino is Dirac or Majorana type.
If the neutrino is Majorana type, the branching ratio is given by
1/15=6.7 \%, while the neutrino is Dirac type, it is given by
2/16=12.5 \%. 
 Therefore, in future, when the data of the direct production of 
$A_i^{\ j}$ are accumulated, we will be able to conclude 
whether neutrinos are Dirac or Majorana by observing whether 
$Br(A_i^{\ j} \rightarrow \nu_i \bar{\nu}_j)$ is $6.7\%$ or $12.5\%$. 




\section{The biggest obstacle against such an optimistic view}


\vspace{2mm}
\noindent
{\bf 3.1 $K^0$-$\bar{K}^0$ mixing }

In the K-Y model, effective current-current interactions with $\Delta N_{fam} =2$            
are given by
$$
H^{eff} = \frac{1}{2} g_F^2 \left[ \sum_i \frac{ (\lambda_i)^2 }{M_{ii}^2} 
+ 2 \sum_{i<j} \frac{\lambda_i \lambda_j }{M_{ij}^2} \right]
(\bar{q}_k \gamma_\mu q_l) (\bar{q}_k \gamma^\mu q_l ) ,
\eqno(3)
$$
where $\lambda_1 = U^*_{1k} U_{1l}$, $\lambda_2 = U^*_{2k} U_{2l}$, 
and $\lambda_3 = U^*_{3k} U_{3l}$.  
For example, for a case of $K^0$-$\bar{K}^0$ mixing are given by
 $\lambda_1=U^{d*}_{11} U^d_{12}$, $\lambda_2= U^{d*}_{21} U^d_{22}$ 
and $\lambda_3= U^{d*}_{31} U^d_{32}$.
These $\lambda_i$ with $k\neq l$ satisfy a unitary triangle condition
$$
\lambda_1 +\lambda_2 + \lambda_3 = 0.
\eqno(4)
$$
We define the effective coupling constant $G^{eff}$ in the current-current
interaction as 
$$
G^{eff} = \frac{1}{2} g_F^2 \left[ \frac{\lambda_1^2}{M_{11}^2} +
 \frac{\lambda_2^2}{M_{22}^2} +  \frac{\lambda_3^2}{M_{33}^2} +
 2\left( \frac{\lambda_1 \lambda_2 }{M_{12}^2} +
 \frac{\lambda_2 \lambda_3 }{M_{23}^2} + \frac{\lambda_3 \lambda_1 }{M_{31}^2}
 \right) \right] .
\eqno(5)
$$

Then, we can estimate the value of $M_{22}$ from the observed value of 
$K^0$-$\bar{K}^0$ mixing by using the expression (5) together with the 
triangle relation (4), i.e. $G^{eff} \simeq (g_f^2/2) \lambda_2^2/M_{22}^2$. 
Under the vacuum-insertion approximation, we obtain 
$\tilde{M}_{22} \sim 340$ TeV from $U^d \simeq V_{CKM}$, the present data 
$\Delta m_K^{obs} = (3.484 \pm 0.006) \times 10^{-18}$ TeV 
and a standard model estimate $\Delta m_K^{SM} \sim 2\times 10^{-18}$ TeV. 
 
\vspace{2mm}
\noindent
{\bf 3.2  Can we build a model with visible gauge bosons?}

In the inverted mass hierarchy model (K-Y model), 
the family gauge boson masses are generated by a scalar with
$({\bf 3},{\bf 3}^*)$ of U(3)$\times$U(3)$'$, so that the 
masses $M_{ij}$ satisfy a relation $2 M_{ij}^2 = M_{ii}^2 + M_{jj}^2$.
Therefore, the mass spectrum is given by 
$$
M_{33} : M_{23} : M_{22} : M_{31} : M_{12} : M_{11} =
1 : \sqrt{ \frac{1+a^2}{2}} : a :  \sqrt{ \frac{1+b^2}{2}} :  
 \sqrt{ \frac{a^2+b^2}{2}} :  b ,
\eqno(6)
$$
where the parameters 
$a$ and $b$ are defined by $a \equiv {M_{22}}/{M_{33}}$ and  
$b \equiv {M_{11}}/{M_{33}}$. 

Regrettably, 
we find that, as far as we consider $\tilde{M}_{22} \sim 340$ TeV, 
we cannot obtain visible values of $\tilde{M}_{33}$ and $\tilde{M}_{12}$, 
although the value $\tilde{M}_{22} \sim 340$ TeV is considerably small 
compared with that from the conventional family gauge models.

\section{An example of a visible family gauge boson model}

Let us demonstrate a toy model. 
We keep the mass relation (6), 
but we regard $a$ and $b$ as free parameters. 
We suppose $b \sim a$.
Note that in the limit of $b=a$ , $G^{eff}_{K\bar{K}}$                
(and also $\Delta m_K$) becomes negligibly small because of  
the observed values $\lambda_3 \simeq 0$ and 
$\lambda_1 \simeq -\lambda_2$.  
Therefore, we can build a visible family gauge boson model without
 using the value $\tilde{M}_{22}\sim 340$ TeV from the 
$K^0$-$\bar{K}^0$ mixing.
 
We still consider that gauge boson mass ratios are deeply related to 
the charged lepton mass rations. 
We start from a world in which  
$(M_{33}^0)^2: (M_{22}^0)^2 : (M_{11}^0)^2 = m_e : m_\mu : m_\tau$,
i.e. $a_0=\sqrt{m_\mu/m_e}$ and $b_0 =\sqrt{m_\tau/m_e}$.   
Next, we suppose a world with $a = b$. 
We speculate that the parameter $a=b$ is given by 
$$
a = b = \sqrt{a_0 b_0} = \left( \frac{m_\mu}{m_e} \frac{m_\tau}{m_e}
\right)^{1/4} = 29.1 .
\eqno(7)
$$
Then, by using the input value $\tilde{M}_{12} \sim 250$ TeV
form the observed $K^+ \rightarrow \pi^+ + \nu +\bar{\nu}$,  we obtain
$$
\tilde{M}_{33} \sim 8.6 \ {\rm TeV}, \ \ 
\tilde{M}_{23} \simeq \tilde{M}_{31}  \sim 177 \ {\rm TeV}, \ \ 
\tilde{M}_{22} \simeq \tilde{M}_{12}  \simeq \tilde{M}_{11} \sim 250 \ {\rm TeV}.
\eqno(8)
$$ 
Therefore, by using $g_F/\sqrt{2} = 0.49742$ which comes from Sumino's 
cancellation condition, we obtain family gauge boson masses:
$$
{M}_{33} \sim 4.3 \ {\rm TeV}, \ \ 
{M}_{23} \simeq {M}_{31}  \sim 88 \ {\rm TeV}, \ \ 
{M}_{22} \simeq {M}_{12}  \simeq {M}_{11} \sim 124 \ {\rm TeV}.
\eqno(9)
$$ 
The result (9) makes the gauge bosons $A_3^{\ 3}$ 
and  $A_1^{\ 2}$ possible to observe at the LHC and at the COMET, 
respectively.

However, this toy model is somewhat unsightly.
We must search for a more natural and  reasonable model.



{\small
\begin{thebibliography}{99}

\bibitem{Sumino_PLB09}
Y.~Sumino, PLB 671, 477 (2009).
%
\bibitem{K-Y_PLB12}
Y.~Koide an d T.Yamashita, PLB 711, 384 (2012).
%
\bibitem{Koide-mass}
  Y.~Koide,
  Lett.\ Nuovo Cim.\  {\bf 34}, 201 (1982);
%
  Phys.\ Lett.\  B {\bf 120}, 161 (1983).
%
%
\bibitem{PDG12}
 J.~Beringer {\it et al.},  Particle Data Group,
 Phys.~Rev.\ D {\bf 86}, 0100001 (2012). 
%
\bibitem{B-Kpinunu}
 G.~Ishidori, F.~Mescia and C.~Smith, Nucl.~Phys. {\bf B 718}, 319 (2005).
%
\bibitem{harmless} 
Y.~Koide, arXiv:1311.5320 [hep-ph]. 
%
\bibitem{COMET}
Y.~Kuno, Prog.~Theor.~Exp.~Phys. 022C01 (2013).
%
%

\end{thebibliography}
}
\end{document}